\documentclass[aps,prc,reprint,twocolumn,superscriptaddress,nofootinbib]{revtex4-2}

\usepackage{bm}
\usepackage{amsmath}
\usepackage{amssymb}
\usepackage{color}
\usepackage{epsfig}
\usepackage[colorlinks,breaklinks]{hyperref}
\usepackage[titletoc]{appendix}
\usepackage{bbm}
\usepackage{multirow}
\newcommand{\be}{\begin{equation}}
\newcommand{\ee}{\end{equation}}

\newcommand{\bes}{\begin{split}}
\newcommand{\ees}{\end{split}}
\newcommand{\ber}{\begin{eqnarray}}
\newcommand{\eer}{\end{eqnarray}}

\newcommand{\bra}{{\langle}}
\newcommand{\ket}{{\rangle}}

\newcommand{\hw}{\ensuremath{\hbar\Omega}}

\newcommand{\SU}[1]{\ensuremath{\mathrm{SU}( #1 )}}
\newcommand{\Un}[1]{\ensuremath{\mathrm{U}( #1 )}}
\newcommand{\SO}[1]{\ensuremath{\mathrm{SO}( #1 )}}

\newcommand{\SpR}[1]{\ensuremath{\mathrm{Sp}( #1,\mathbb{R} )}}

\usepackage{capt-of}
\usepackage{relsize}
\DeclareMathOperator*{\SumInt}{%
	\mathchoice%
	{\ooalign{$\displaystyle\sum$\cr\hidewidth$\displaystyle\int$\hidewidth\cr}}
	{\ooalign{\raisebox{.14\height}{\scalebox{.7}{$\textstyle\sum$}}\cr\hidewidth$\textstyle\int$\hidewidth\cr}}   
	{\ooalign{\raisebox{.2\height}{\scalebox{.6}{$\scriptstyle\sum$}}\cr$\scriptstyle\int$\cr}}
	{\ooalign{\raisebox{.2\height}{\scalebox{.6}{$\scriptstyle\sum$}}\cr$\scriptstyle\int$\cr}}  
}

\begin{document}

\title {Response functions and giant monopole resonances for light to medium-mass nuclei from the \textit{ab initio} symmetry-adapted no-core shell model}

\author{M.~Burrows} \affiliation{Department of Physics and Astronomy, Louisiana State University, Baton Rouge, LA 70803, USA}

\author{R.~B. Baker} \affiliation{Department of Physics and Astronomy, Louisiana State University, Baton Rouge, LA 70803, USA}
\affiliation{Institute of Nuclear and Particle Physics, and Department of Physics and Astronomy, Ohio University, Athens, Ohio 45701, USA}

\author {S. Bacca} \affiliation{Institut f\"ur Kernphysik and PRISMA$^+$ Cluster of Excellence, Johannes Gutenberg-Universit\"at Mainz, 55128 Mainz, Germany} \affiliation{Helmholtz Institute Mainz, (Germany), GSI Helmholtzzentrum f\"ur Schwerionenforschung, Darmstadt, Germany}

\author{K.~D. Launey} \affiliation{Department of Physics and Astronomy, Louisiana State University, Baton Rouge, LA 70803, USA}

\author{T. Dytrych}
\affiliation{Nuclear Physics Institute, Academy of Sciences of the Czech Republic, 250 68 Řež, Czech Republic}
\affiliation{Department of Physics and Astronomy, Louisiana State University, Baton Rouge, LA 70803, USA}

\author{D. Langr}
\affiliation{Department of Computer Systems, Faculty of Information Technology, Czech Technical University in Prague, Prague 16000, Czech Republic}

\date{\today}

\begin{abstract}
Using the \textit{ab initio} symmetry-adapted no-core shell model, we compute sum rules and response functions for light to medium-mass nuclei, starting from interactions that are derived in the chiral effective field theory. We investigate electromagnetic transitions of monopole, dipole and quadrupole nature for symmetric nuclei such as $^4$He, $^{16}$O, $^{20}$Ne and $^{40}$Ca. Furthermore, we study giant monopole resonance, which can provide information on the incompressibility of symmetric nuclear matter.

\end{abstract}

\maketitle

\section{Introduction}
\label{sec:intro}

Giant resonances are broad resonances that exhaust a major portion of the corresponding non-energy weighted sum rule, the zeroth moment of the nuclear response to an external electromagnetic probe \cite{Harakeh2001}. These have long been studied via the random phase approximation, primarily through the use of schematic or phenomenological interactions \cite{BohrMot2, Goeke1982}. In addition, giant monopole (GMR) and quadrupole (GQR) resonances have been naturally described in the algebraic symplectic \SpR{3} shell model \cite{BahriR00,Rowe16}, as well as  in the no-core symplectic shell model \cite{DreyfussLTDBDB16}, since the corresponding monopole and quadrupole operators do not mix \SpR{3}-symmetry preserving subspaces. While \textit{ab initio} methods have recently started to successfully describe giant resonances~\cite{Bacca:2013_PRL,Miorelli2018,Bacca_2018_JPCS,Stumpf2017,Miorelli2016}, continued  first-principles studies can help elucidate the underpinning microscopic physics and provide an important probe of the underlying realistic interactions. Indeed, due to the perturbative nature of the electromagnetic interaction, calculations of electromagnetic transitions and responses in atomic nuclei can be compared in a straightforward way to experimental data and important features of the strongly interacting nuclear system can be studied~\cite{BaccaPastore2014}. Furthermore, certain giant resonances have connections to other branches of physics, such as the giant monopole resonance (also called ``breathing mode") that can be related to the (in)compressibility of  nuclear matter. Nuclear compressibility is one of the main ingredients to the nuclear equation of state and its astrophysical applications span from core-collapse supernovae to neutron stars \cite{Garg2018}.

Considerable progress has been achieved in computing response functions and sum rules with \textit{ab initio} approaches that describe the nucleus as a system of protons and neutrons interacting with each other as well as with external probes, and solve the problem exactly or with controlled approximations~\cite{Bacca:2014tla}. 

 Response functions and sum rules have been successfully calculated in the shell model \cite{LuJohnson_PRC2018} or using  \textit{ab initio} methods, such as hyperspherical harmonics (HH) and no-core shell model (NCSM) for light nuclei \cite{Quaglioni2007370, 0954-3899-41-12-123002, Stetcu:2007_NPA,BakerLBND20} or the coupled-cluster (CC) method thus far for closed-shell light and medium-mass nuclei \cite{Bacca:2013_PRL, PhysRevC.90.064619, Bacca_2018_JPCS}.

To this end, the symmetry-adapted no-core shell model
 (SA-NCSM)~\cite{DytrychSBDV_PRL07,DytrychSDBV08_review,LauneyDD16,DytrychLDRWRBB20} 
has been shown to be a valuable approach capable of using only physically relevant model spaces with dimensions that are only a fraction of the standard NCSM model space, thereby extending the no-core shell-model reach toward heavier nuclei while maintaining important physical features, such as collectivity and clustering. 

The main purpose of this work is to utilize wave functions calculated in the \textit{ab initio} SA-NCSM for light up through medium-mass nuclei with the goal to study response functions and to probe the nuclear compressibility predicted by realistic interactions without renormalization in the nuclear medium. 

We show that the SA-NCSM can be applied to calculate response functions for intermediate-mass open-shell nuclei and medium-mass nuclei, specifically, for the illustrative examples of $^{20}$Ne and $^{40}$Ca.
The calculations are based on  the Lanczos sum rule method (LSR)~\cite{NND_PRC_2014}, with a recent benchmark study \cite{Baker2020} of $^4$He compared to the  exact solutions in the HH method using chiral nucleon-nucleon (NN) potentials. In the benchmark study of Ref. \cite{Baker2020}, the SA-NCSM results have been calculated in selected model spaces and compared against those in the corresponding complete model spaces, which have recovered the outcomes of the standard NCSM \cite{NavratilVB00,BarrettNV13}. Indeed, we have shown good agreement, while using much smaller model spaces, corroborating earlier finding for structure observables and form factors \cite{DytrychHLDMVLO14,DytrychMLDVCLCS11,LauneyDD16}. Here, we expand the analysis to heavier nuclei systems, with a focus on monopole response functions and giant monopole resonances.

 This paper is organized as follows. In Section \ref{sec:methods} we provide a brief overview of the  methods used in this study, namely the Lanczos sum rule/response method and the symmetry-adapted no-core shell model. In Section \ref{sec:res}, we present results for the various electromagnetic sum rules and response functions considered here, investigate properties of giant monopole resonances, and discuss the case of the nuclear compressibility.  Finally, in Section \ref{sec:conclude} we present our conclusions.

\section{Theoretical framework}
\label{sec:methods}

\subsection{Lanczos method for sum rules and responses }

The response of a nucleus
to an external perturbation of energy $E_{\rm X}$ is described 
by the response function, defined as
\begin{equation}\label{eq:S_omega}
R(E_{\rm x})=\SumInt_f |\bra \psi_f |\hat{O}| \psi_0 \ket|^2 \delta\left(E_f-E_0-E_{\rm x}\right),
\end{equation}
where $\hat{O}$ is the operator that induces a transition from the initial state $| \psi_0\ket$ into a set of final states $ |\psi_f \ket$. Here, 
$E_{0(f)}$ 
are the eigenvalues
of the Hamiltonian $\hat{H}$ 
for the initial and final state, respectively,
while the symbol $\SumInt_f$ indicates the inclusion of the entire discrete and continuous spectrum, such that $\SumInt_f |\psi_f \ket \bra \psi_f | = \mathbf{1}$. We consider
three electromagnetic operators $\hat O$, relevant to
nuclear structure, namely, the isoscalar 
electric monopole operators (carrying angular momentum $L=0$), the electric dipole operators ($L=1$), and the electric quadrupole operators $(L=2)$,   defined respectively as
\begin{eqnarray}
\hat{M} &=& \frac{1}{2} \sum^A_{i=1} r_i^2 \\
\hat{D} &=& \sqrt{\frac{4\pi}{3}} \sum^A_{i=1} e_i r_i Y_{10}(\hat{r}_i) \label{eqn:dip_op} \\
\hat{Q} &=& \sqrt{\frac{16\pi}{5}} \sum^A_{i=1} e_i r_i^2 Y_{20}(\hat{r}_i),
\end{eqnarray}
where $e_i$ and $\vec r_i$ denote the charge and coordinates of the $i$-th particle.  These coordinates $\vec r_i$, in the no-core shell-model framework, are particle coordinates in the laboratory frame, and hence, the operators in Eq.~(\ref{eqn:dip_op}) are not translationally invariant. Consequently, special care is taken to remove the resulting spurious  CM contribution to the SA-NCSM response function, as detailed in Ref. \cite{BakerLBND20}.

In this work, we focus on several moments of the response function, so-called sum rules, of the form
\begin{equation}\label{eq:SR_mom_integral}
m_n = \int \, dE_{\rm x} \, R(E_{\rm x}) \, E_{\rm x}^n,
\end{equation}
which, using the completeness of the eigenstates $|\psi_f \ket$, can be rewritten as
\begin{equation}\label{eq:SR_mom_braket}
m_n = \bra \psi_0 | \hat{O}^\dagger\, \left( \hat H - E_0\right)^n\, \hat{O} | \psi_0\ket,
\end{equation}
with $m_0=\bra \psi_0 | \hat{O}^\dagger  \hat{O} | \psi_0\ket$ being the non-energy weighted sum rule (NEWSR) or the total strength of the response function. We also discuss $m_1$ and $m_{-1}$, which are called the energy weighted sum rule (EWSR) and inverse energy weighted sum rule (IEWSR), respectively.

According to Eq. (\ref{eq:SR_mom_braket}), the calculation of $m_n$ does not require explicit knowledge of the
excited states in the continuum.
In fact, if the initial state $|\psi_0 \ket$
is localized
and well described within the range of the interaction, 
 which is the case for the ground state of a bound nucleus, then one needs to calculate an expectation value of a many-body operator in the $|\psi_0 \ket$ state.
 However, in practice, one uses the completeness relation in terms of the eigenstates of $\hat{H}$, and only needs the matrix elements of the one-body excitation operator, while the Hamiltonian will obviously be diagonal.  By truncating the  Hilbert space until convergence is reached one can retrieve $m_n$.
In this procedure, it is perfectly justified to use a bound-state method to calculate the excited states that are entering the completeness~\cite{NND_PRC_2014}.

When direct diagonalization of the Hamiltonian matrix is computationally impractical, one can use the Lanczos algorithm instead and apply it directly on the sum rules calculations, resulting in the so called Lanczos sum rule (LSR) method (see, e.g., \cite{Dagotto:1994_RMP,NND_PRC_2014} and references therein).
 This approach leads to
\begin{eqnarray} \label{eqn:LSR_eq}
m_n = \bra \psi_0 | \hat{O}^{\dagger} \hat{O}|\psi_0 \ket \sum_{k=0}^{N_L-1} |Q_{k 0}|^2 (E_{{\rm x},k})^n,
\end{eqnarray}
where $N_L$ is the number of Lanczos iterations, $Q_{k 0}$ is the matrix that diagonalizes the tridiagonal Lanczos matrix, $E_{x,k}$ is the excitation energy of the $k$-th state, and the Lanczos pivot (the starting vector of the iterative tridiagonalization process) is the normalized state $|\phi_0\ket =  \hat{O}|\psi_0 \ket /\sqrt{m_0} $.
The LSR method has been shown to be very efficacious \cite{NND_PRC_2014} and has, for example, allowed to reach the required precision in the calculations of  nuclear structure corrections to the Lamb shift of light muonic atoms~\cite{Ji13,Nevo14,NND:2016_PhysLettB,Ji_2018}.
Furthermore, the method has been recently applied to calculations based on coupled-cluster theory~\cite{Miorelli2016,Hagen16,Miorelli2018,PhysRevLett.124.132502,PhysRevC.102.064312,PhysRevC.105.034313,PhysRevResearch.5.L022044} and SA-NCSM~\cite{BakerLBND20}. 

Response functions can be obtained without explicitly solving for the final eigenstates with continuum boundary conditions by utilizing integral transform methods. A prominent example is the Lorentz integral transform (LIT), which has been well documented in the literature and used to obtain nuclear responses for electromagnetic and weak operators~\cite{Efros:1994_PLB,Efros:2007_JPG,Doron}. 
The LIT is defined as
\begin{eqnarray}
\label{lit}
\mathcal{L}(\sigma_p,\Gamma) = \frac{\Gamma}{\pi} \int dE_{\rm x} \frac{R(E_{\rm x})}{(E_{\rm x} - \sigma_p)^2 + \Gamma^2},
\end{eqnarray}
where $\sigma_p$ and $\Gamma$ determine the peak position and width of the Lorentzian kernel, respectively. In this method one typically calculates the LIT for a set of values of the parameters  $\sigma_p$ and $\Gamma$ using a bound state method and then one retrieves the response function  by performing a regularized inversion of the integral transform, see Ref.~\cite{Efros:2007_JPG}. In this way one recovers an $R(E_{\rm x})$ which is (almost) independent on the resolution scale $\Gamma$.

One may proceed in a different way by using Eq.~(\ref{eq:S_omega}) and computing the discretized response function  for discrete eigenstates $|\psi_f \ket$, and then folding the strength function of each discrete state with a Lorentzian function of width $\Gamma$, noting that $\lim_{\Gamma \rightarrow 0} \frac{\Gamma}{\pi} \frac{1}{(E_{\rm x}-E_{f}+E_0)^2 + \Gamma^2}= \delta(E_{\rm x}-E_{f}+E_0)$. The folding procedure has been extensively used in mean field approaches, see e.g.~\cite{Paar_2007} and references therein. 
When the direct diagonalization is impractical due to the large size of the matrices, one may use the Lanczos method to tridiagonalize $\hat{H}$, together with the expression analogous to Eq.~(\ref{eqn:LSR_eq}), sometimes called the Lanczos response method \cite{Efros:2007_JPG, BakerSOTANCP42018}:
\begin{eqnarray} \label{eqn:LSR_lit}
&&R_n(E_{\rm x},\Gamma) \rightarrow  \mathcal{L}_n(E_{\rm x},\Gamma)=\\
\nonumber
&&=\bra \psi_0 | \hat{O}^{\dagger} \hat{O}|\psi_0 \ket \sum_{k=0}^{N_L-1} |Q_{k 0}|^2 \frac{\Gamma}{\pi} \frac{(E_{{\rm x},k})^n}{(E_{\rm x}-E_{{\rm x},k})^2 + \Gamma^2}.
\end{eqnarray}
When computing responses functions, this is essentially equivalent to calculating Eq.~(\ref{lit})  for a finite and small value of $\Gamma$~\cite{Efros:2007_JPG} and interpreting  $\mathcal{L}(\sigma_p,\Gamma)$ as the response function   $R(E_{\rm x})$ itself that now depends on the resolution scale $\Gamma$.  
This is the approach that we will follow in this paper.
In this way, one does not perform the delicate inversion procedure, yet one can  study properties of the response functions, including fragmentations and giant resonances, albeit with an explicit resolution-scale dependence.

As a bound-state method we will employ the SA-NCSM, which provides us with high-quality wave functions, which can be decomposed and examined in terms of individual basis states and their associated deformation.

\subsection{Symmetry-adapted no-core shell model}

\begin{figure}[t]
\centering
\includegraphics[width=0.3\textwidth]{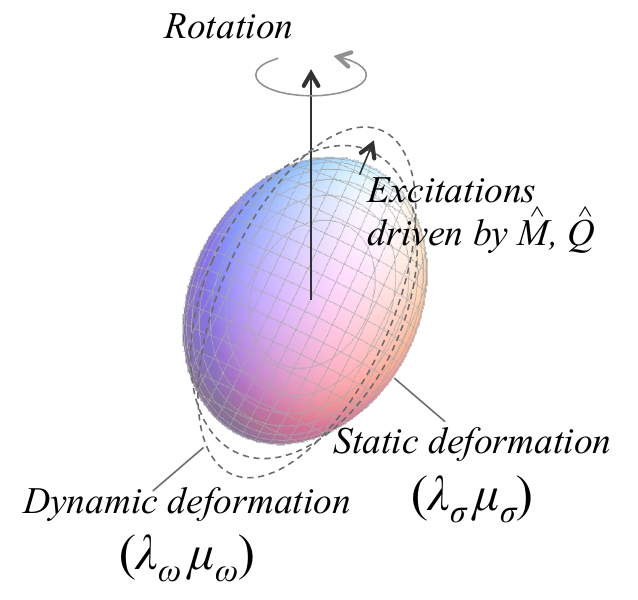} 
\caption{Nuclear shape consisting of a static (equilibrium) deformation $(\lambda_\sigma\,\mu_\sigma)$ (bandhead) and dynamic deformations $(\lambda_\omega\,\mu_\omega)$ (vibrations) that are multiples of $2\hw$ one-particle-one-hole excitations driven by the monopole $\hat M$ and quadrupole $\hat Q$ operators, along with their rotations. A nuclear shape belongs to an \SpR{3}-preserving subspace, labeled by $\sigma=N_\sigma(\lambda_\sigma\,\mu_\sigma)$ whereas a deformation belongs to an \SU{3} preserving subspace labeled by $\omega=N(\lambda_\omega\,\mu_\omega)$ with $N=N_x+N_y+N_z \leq N_{\rm max}$, $\lambda_\omega = N_z-N_x$ and $\mu_\omega=N_x-N_y$ for $N_z$, $N_x$, and $N_y$ total HO excitations in the $z$, $x$, and $y$ directions, respectively. }
\label{fig:shape}
\end{figure}

The SA-NCSM framework \cite{DytrychSBDV_PRL07,DytrychLDRWRBB20} (reviewed in \cite{LauneyDD16,LauneyMD_ARNPS21}) is an {\it ab initio} no-core shell model that employs a symmetry-adapted basis, an \SU{3}$\supset$\SO{3}-coupled basis or an \SpR{3}$\supset$\SU{3}$\supset$\SO{3}-coupled basis with an SA selection based on the \SpR{3} symmetry.
The significance of the \SU{3} group for a
microscopic description of the nuclear dynamics can be seen from the fact that it is the symmetry group of the harmonic oscillator (HO) utilized in the successful Elliott model~\cite{Elliott58,Elliott58b}, and a subgroup of the physically relevant \SpR{3} symplectic model~\cite{RosensteelR77, Rowe85,Rowe96}, which provides a comprehensive microscopic foundation for understanding the dominant symmetries of nuclear dynamics.
The symmetry-adapted (SA) concept is based on the idea that the infinite Hilbert space can be equivalently spanned by ``microscopic" nuclear shapes and their rotations, where  ``microscopic" means that these configurations track with position and momentum coordinates of each particle. A nuclear shape belongs entirely to a  single symplectic-preserving subspace called ``irrep"; it can be viewed as  a ``static" (equilibrium) deformation and its vibrations of the giant-resonance (GR) type, driven by the monopole $\hat M$ and quadrupole $\hat Q$ operators, that are  understood as  ``dynamical" deformations  (Fig. \ref{fig:shape}). An important advantage of the SA-NCSM is that many spherical and less deformed shapes and their mixing are already included in typical small shell-model spaces, however, the vibrations of largely deformed static deformations and  spatially extended modes like clustering often lie outside such spaces, but are included in the SA model spaces. 

To calculate response functions in this work, we use \Un{3}$\supset$\SU{3}$\supset$\SO{3} to label the basis states, together with $S_{p}$, $S_{n}$, and $S$ that denote proton, neutron, and total intrinsic spins. 
An \SU{3} irrep is labeled by a set of quantum numbers $(\lambda_\omega\,\mu_\omega)$ (see Fig. \ref{fig:shape}). They bring forward important information about nuclear shapes and
deformation, according to an established mapping \cite{CastanosDL88,RosensteelR77,LeschberD87}; for example, $(0 0)$,
$(8\, 0)$ and $(0\,8)$ describe spherical, prolate and oblate deformation, respectively.

As in the NCSM, the particle coordinates are specified in the laboratory frame and we use the Lawson technique to exactly treat the removal of the spurious center of mass motion \cite{Lawson74,Verhaar60,Hecht71,Millener92,LauneyDD16, BakerLBND20}. We employ the many-body $N_{\rm max}$ truncation where the $N_{\rm max}$  cutoff is defined as the maximum number of HO quanta allowed in a many-particle state above the minimum for a given nucleus. Hence, basis states where one nucleon carries all the $N_{\max}$ quanta are included, in which cases one nucleon occupies the highest HO shell.
In the SA-NCSM, we adopt a notation where an SA-NCSM model space of ``$\langle N_0 \rangle N_{\max}$'' includes all the basis states up through $N_0$  total excitations and a selected basis states in  $N_0+2$, $N_0+4$,... up through $N_{\max}$. The selection is based on a robust prescription outlined in Ref. \cite{LauneyDSBD20}: one solves the many-body eigenproblem typically in $N_0$ or $N_0+2$ complete model space, identifies the nonnegligible configurations within the nuclear eigenstate, and then constructs the selection upon these configurations by symplectic \SpR{3} excitations thereof up to large $N_{\max}$ values that are inaccessible to complete-space computations. Configurations of the largest deformations (typically, large $\lambda$ and $\mu$) and lowest spin values are included first. 
\begin{figure*}[th]
\centering
\begin{tabular}{ccc}
(a) $L=0$	& 	(b)	$L=1$& 	(c)	$L=2$\\
\includegraphics[width=0.33\textwidth]{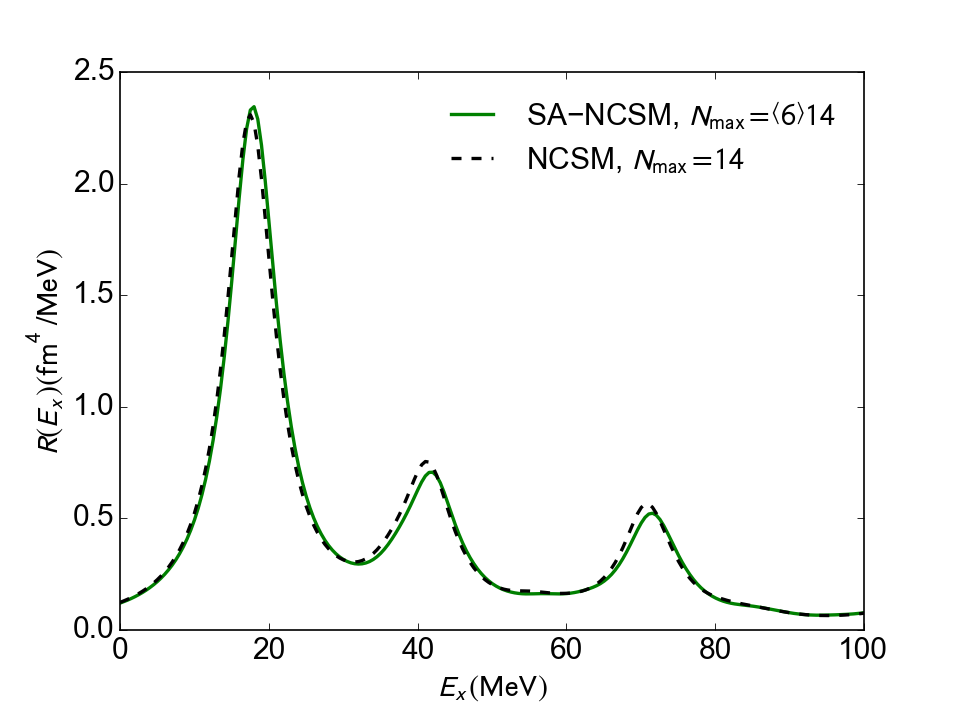} & 
\includegraphics[width=0.33\textwidth]{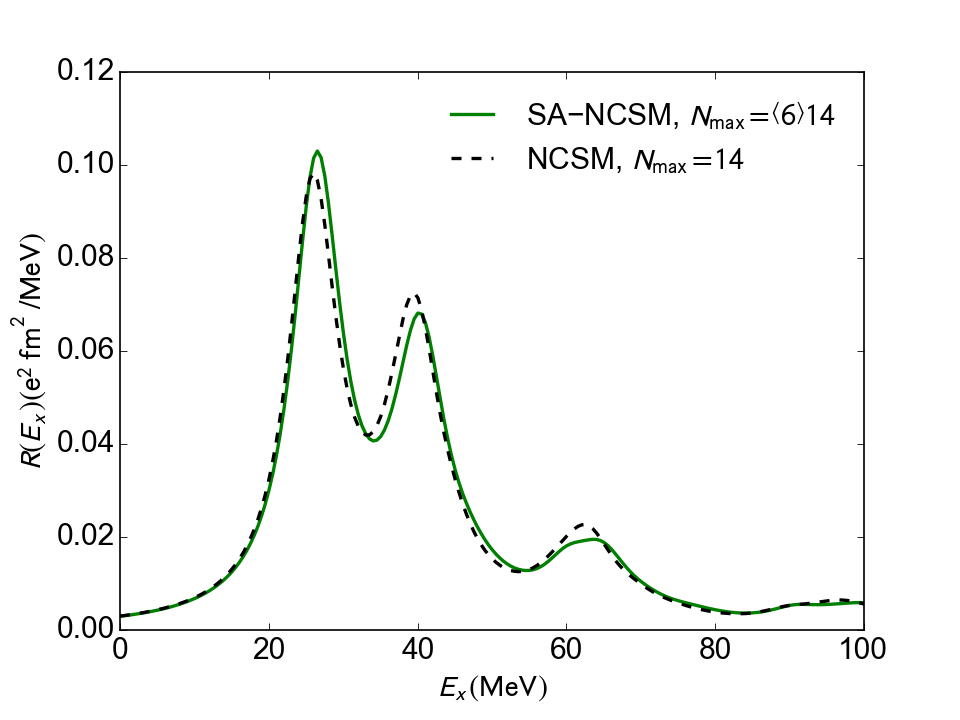}	&	
\includegraphics[width=0.33\textwidth]{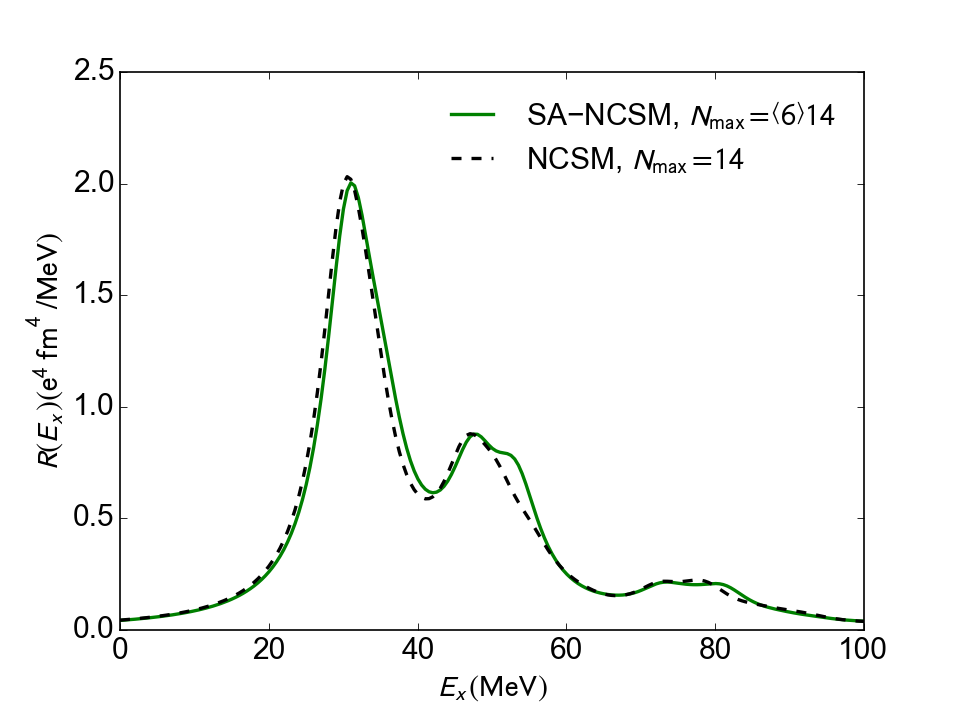}
\end{tabular}
\caption[\protect\vspace{-2ex}{Monopole, dipole, and quadrupole response functions for $^4$He from the NCSM and SA-NCSM with NNLO$_{\mathrm{opt}}$}]{(a) Monopole, (b) dipole, and (c) quadrupole response functions for $^4$He from the NCSM and SA-NCSM with NNLO$_{\mathrm{opt}}$ in $N_{\mathrm{max}}=14$ and $N_{\mathrm{max}}=\bra 6 \ket 14$ model spaces, respectively. All of the response functions are shown for a width of $\Gamma=4$ MeV and $\hbar\Omega=25$ MeV.}
\label{fig:He-4_resp}
\end{figure*}

\section{Results and Discussions}
\label{sec:res}

We present SA-NCSM calculations for the light nucleus of $^4$He, spherical $^{16}$O, deformed $^{20}$Ne, and medium-mass $^{40}$Ca, using chiral effective field theory (EFT) potentials without renormalization in nuclear medium.  In particular, we adopt the NNLO$_{\rm opt}$  chiral nucleon-nucleon (NN) potential \cite{Ekstrom13}, as well as  NNLO$_\mathrm{sat}$ \cite{PhysRevC.91.051301} with the three-nucleon (NNN) forces, hierarchically smaller than their NN forces, added as averages and denoted as NNN(0) \cite{LauneyMD_ARNPS21}. The NNLO$_{\rm opt}$ is used without NNN forces, which have been shown to contribute minimally to the 3- and 4-nucleon binding energy \cite{Ekstrom13}. Remarkably, the NNLO$_{\rm opt}$ NN potential has been found to reproduce various observables and to agree with the outcomes of chiral NN+NNN potentials, including, e.g., the $^4$He electric dipole polarizability \cite{BakerLBND20}; the challenging analyzing power for elastic proton scattering on $^4$He, $^{12}$C, and $^{16}$O \cite{BurrowsEWLMNP19}; neutron-deuteron scattering cross sections \cite{PhysRevC.106.024001}; along with  B(E2) transition strengths  for $^{21}$Mg and $^{21}$F  \cite{Ruotsalainen19} in the SA-NCSM without effective charges.
\begin{figure*}[th]
\centering
\begin{tabular}{ccc}
(a) $^{16}$O	& 	(b) $^{20}$Ne	& 	(c)	$^{40}$Ca \\
\includegraphics[width=0.33\textwidth]{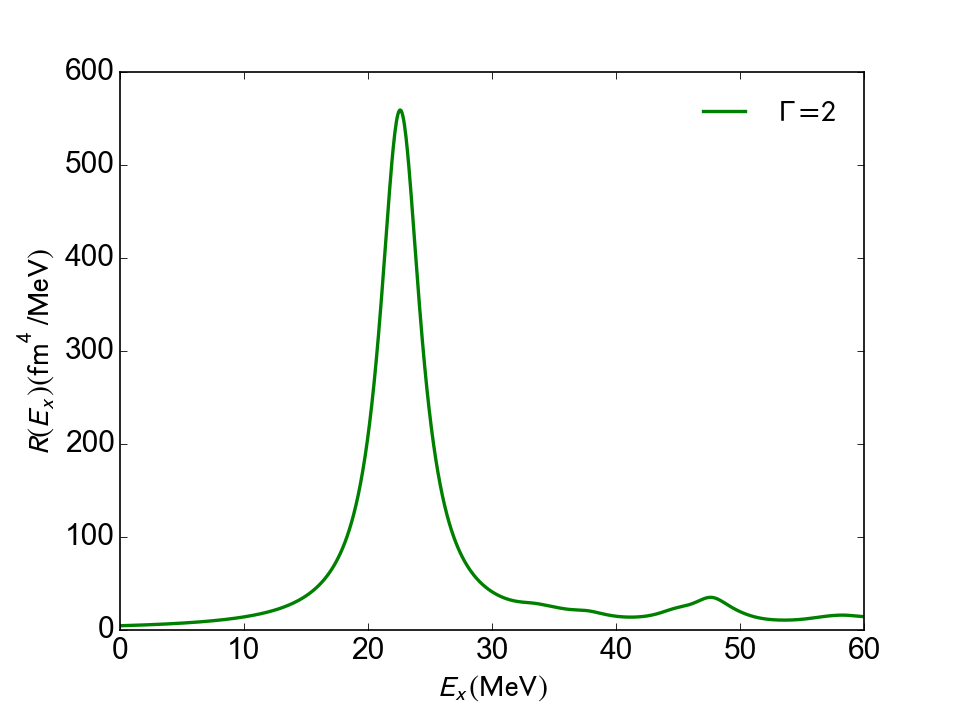} & 
\includegraphics[width=0.33\textwidth]{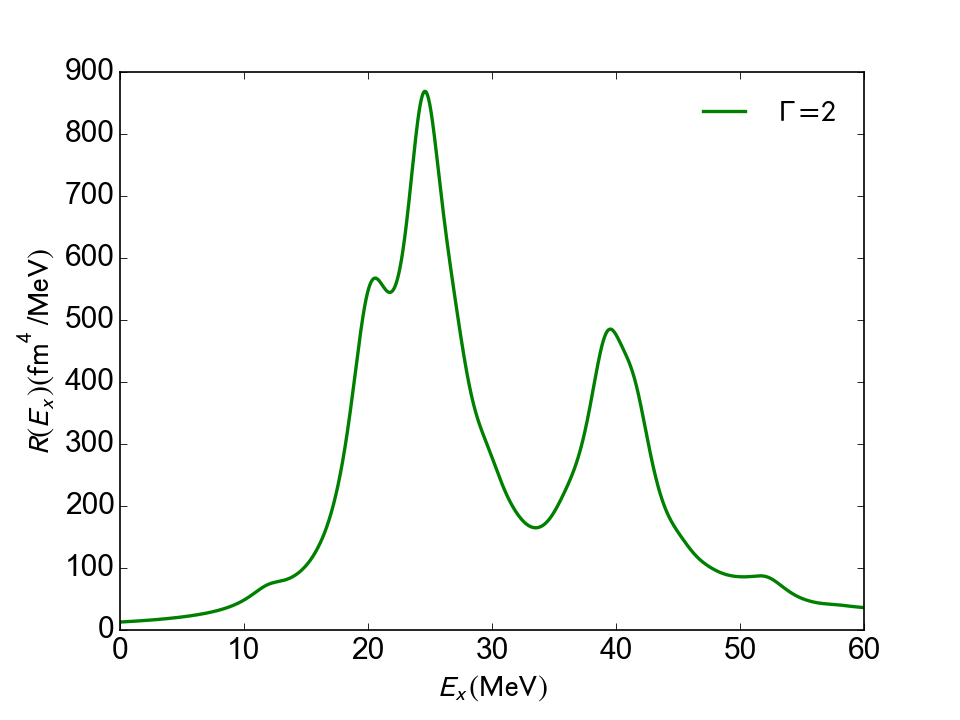}	&	
\includegraphics[width=0.33\textwidth]{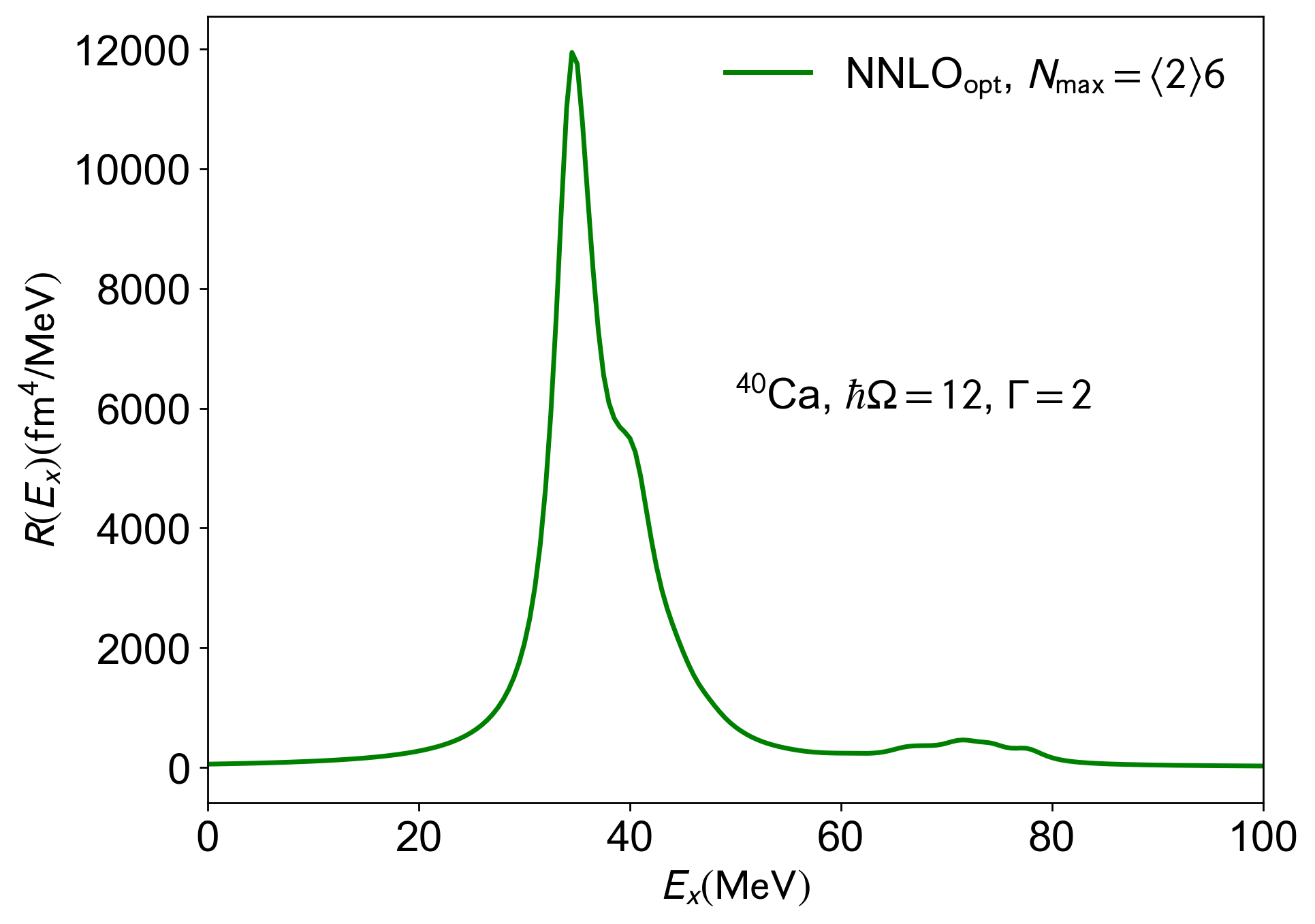}
\end{tabular}
\caption{Monopole response functions in (a) $^{16}$O at $N_{\mathrm{max}}=\bra 2 \ket 10$ and $\hbar\Omega=16$ MeV, (b) $^{20}$Ne at $N_{\mathrm{max}}=\bra 2 \ket 10$ and $\hbar\Omega=15$ MeV, and (c) $^{40}$Ca at $N_{\mathrm{max}}=\bra 2 \ket 6$ and $\hbar\Omega=12$ MeV. All response functions are calculated with NNLO$_{\mathrm{opt}}$ and a width of $\Gamma=2$ MeV.}
\label{fig:mono_resp}
\end{figure*}

We study nuclear responses with a focus on monopole response functions that provide an \textit{ab initio} description of giant monopole resonances (GMRs). The GMRs, in turn, inform incompressibility in nuclei and can be used to guide nuclear matter incompressibility  properties. Indeed, for almost incompressible nuclear matter, the one-phonon monopole excitation, the so-called breathing mode, is expected to be observed at a high excitation energy \cite{Rowe16}. To discuss nuclear features, we present responses for an optimal \hw~value, where the convergence of monopole moments or radii is typically achieved at comparatively smaller model spaces, and which is often given as $\sim 41/A^{1/3}$ \cite{LauneyDD16}. Specifically, we use \hw=25 MeV for $^4$He (whereas a comprehensive analysis of the monopole sum rule for $^4$He is available in our earlier study \cite{BakerLBND20}), as well as \hw=16, 15, and 12 MeV for $^{16}$O, $^{20}$Ne, and $^{40}$Ca, respectively. For sum rules, we report uncertainties for small variations in \hw~around these optimal values.

Using the $m_1$ energy weighted sum rule and the $m_{-1}$ inverse
energy weighted sum rule for monopole transitions (e.g., see Ref. \cite{PhysRevC.70.024307}), we are able to calculate the  GMR centroid  energy $E_{\mathrm{GMR}}$ form the \textit{ab initio} SA-NCSM calculations as 
\begin{eqnarray}
\label{eqn:GMR_SR}
E_{\mathrm{GMR}} = \sqrt{\frac{m_1}{m_{-1}}}.
\end{eqnarray}
This centroid energy includes the fragmentation of the GMR across the entire energy region and in general differs from the energy of the first main peak in the monopole response functions, as discussed in the next section. The peak energy can be also measured by experiments that probe energy regions up through $\sim 40$ MeV. In addition, the LSR method calculates $m_{\pm 1}/m_0$ through the Lanczos coefficients. This means that the $m_{\pm 1}$ sum rule includes errors from the LSR calculations and $m_0$. However, smaller errors can be obtained by using $\sqrt{\frac{m_1/m_0}{m_{-1}/m_0}}$ in Eq. (\ref{eqn:GMR_SR}).

\subsection{Nuclear responses and giant monopole resonances}
\label{sec:resp}
We  first examine the $^4$He response functions for the monopole ($L=0$), dipole ($L=1$), and quadrupole ($L=2$) operators calculated from the Lanczos response method (\ref{eqn:LSR_lit}), using $\Gamma=4$ MeV and NNLO$_{\rm opt}$ (Fig.~\ref{fig:He-4_resp}).  We observe that the response functions peak at 17.8 MeV for the $0^+$ states in $^4$He with nonzero monopole transitions to the ground state, as shown in Fig.~\ref{fig:He-4_resp}a, at 26.5 MeV for $1^-$ states in Fig.~\ref{fig:He-4_resp}b, and at 31.0 MeV for $2^+$ states in Fig.~\ref{fig:He-4_resp}c, and thus these peaks occur closely to the corresponding lowest isospin-zero 20.21-MeV 0$^+$ state, 24.25-MeV 1$^-$ state, and 27.42-MeV 2$^+$ state in the $^4$He experimental energy spectrum. Hence, these 0$^+$  and 2$^+$ states can be understood as giant monopole and quadrupole resonances, respectively, governed by 2\hw~one-particle-one-hole excitations of the ground state (see also Ref. \cite{BaccaBLO13}). Remarkably, since the monopole and quadrupole operators are generators of the symplectic \SpR{3} symmetry and do not mix symmetry-preserving subspaces, the $L=0$ ($L=2$) response functions reach excited $0^+$ ($2^+$) states that necessarily contain the symplectic structure (shapes) of the ground state. It will be interesting to calculate the response functions in a full systematic study for several \hw~values, including uncertainty quantification arising from the many-body approach and the underlying interaction to compare to the recent experimental results on the $\alpha$ monopole transition form factor and the monopole transition matrix element \cite{Kegel:2021jrh}.

Furthermore, the NCSM and SA-NCSM response functions are in good agreement, regardless of the operator. This suggests that we can utilize the SA selection for response functions for heavier nuclei, even in model spaces beyond the reach of the standard NCSM, such as $N_{\rm max}=10$ for $^{20}$Ne (with $7 \times 10^{10}$ dimensionality of the complete $J^\pi=0^+$ model space) and $N_{\rm max}=6$ for $^{40}$Ca (with $0.3 \times 10^9$ dimensionality of the complete $J^\pi=0^+$ model space). 

To report a centroid energy for the monopole distribution in $^4$He, we use  Eq. (\ref{eqn:GMR_SR}) and $m_{1}=176.7(4)$ fm$^4$MeV, $m_{-1}=0.242(2)$ fm$^4$MeV$^{-1}$, and $m_{0}=22.92(4)$ fm$^4$ for the NNLO$_{\rm opt}$ chiral potential. This yields a centroid energy $E_{\mathrm{GMR}}$ for the monopole distribution in $^4$He of $27.0(1)$ MeV (Table \ref{tab:obs}), which is indeed larger than the energy of the first peak in the monopole response function shown in Fig.~\ref{fig:He-4_resp}a.

Examining the monopole response for $^{16}$O, using Eq. (\ref{eqn:LSR_lit}) with $\Gamma=2$ MeV and NNLO$_{\rm opt}$ (Fig.~\ref{fig:mono_resp}a), we find one well-defined peak around $23.5$ MeV, providing strong evidence of the giant monopole resonance. Experimental work on the giant resonances in $^{16}$O have indeed found values consistent with this energy range (see, e.g.~Ref.~\cite{Lui2001} and references therein). For $E_{\mathrm{GMR}}$, Ref.~\cite{Lui2001} reports a GMR centroid of $21.13 \pm 0.49$ MeV, which is very close to our estimate of $24(1)$ MeV (Table \ref{tab:obs}).
In addition, using the ability of the SA-NCSM to determine the intrinsic shape and deformation of nuclear states, we find that this peak  is dominated ($>60\%$) by $\omega=2(2\ 0)$ dynamic deformation of correlated $2\hbar\Omega$ one-particle-one-hole excitations of a spherical equilibrium shape, the same one that dominates the ground state of $^{16}$O. This is inline with previous work that uses the no-core symplectic shell model and an EFT-inspired inter-nucleon interaction \cite{DreyfussLTDBDB16}.
Here, for the first time, we find this feature  emerging from the underlying chiral potential. 

Interestingly, a very similar behavior is observed for the monopole response for the heavier closed-shell $^{40}$Ca nucleus (Fig.~\ref{fig:mono_resp}c). Similar to $^{16}$O, the peak in the monopole resonance function for $^{40}$Ca is dominated by a $\omega=2(2\ 0)$ dynamic deformation, which describes the GR-type vibrations  of the spherical equilibrium shape that dominates the ground state of $^{40}$Ca.

\begin{figure}[t]
\centering
\includegraphics[width=0.45\textwidth]{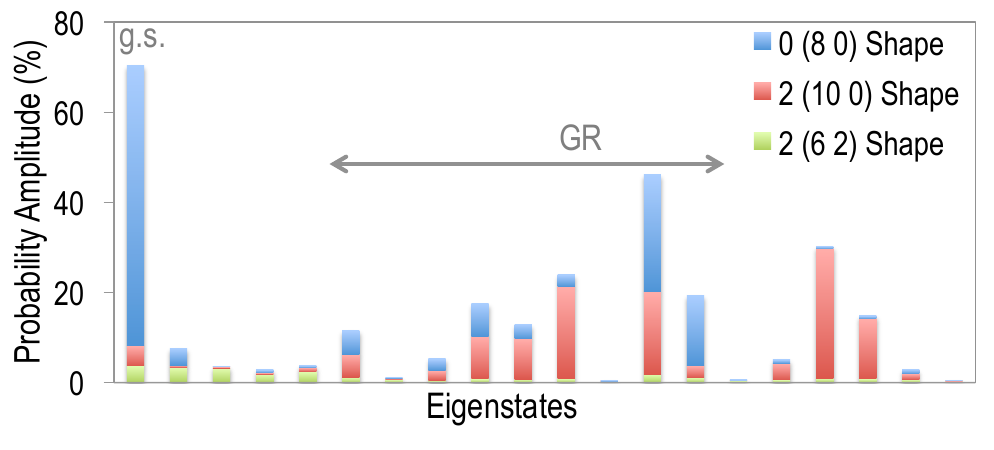}
\caption{Nuclear shape contributions to the 20 lowest $^{20}$Ne eigenstates, calculated with NNLO$_{\mathrm{opt}}$, $\hbar\Omega=15$ MeV, and for 11 HO shells. Eigenstates are ordered in increasing eigenenergy. The three most dominant shapes in the ground state are shown only, namely, the predominant $\sigma=0(8\ 0)$ shape, along with $2(10\ 0)$ and $2(6\ 2)$ shapes.}
\label{fig:Ne20_sp}
\end{figure}

The situation is more interesting for an open-shell deformed nucleus, such as the case of the monopole response for $^{20}$Ne (Fig.~\ref{fig:mono_resp}b). Unlike the $^{16}$O case, the GMR strength is no longer concentrated in a single peak, but is instead fragmented across the energy range $20-40$ MeV. In addition, the first peak at $\sim 25$ MeV contains $2\hbar\Omega$ excitations of the $\sigma=0(8\ 0)$ shape that is known to dominate the ground state (see Fig.~\ref{fig:Ne20_sp} for the three most dominant shapes in the ground state and their contribution to the lowest 19 excited $0^+$ states).  This corroborates the features of a fragmented giant resonance suggested in Ref. \cite{DytrychLDRWRBB20} based on the contribution to the excited $^{20}$Ne $0^+$ states of the $0(8\ 0)$ g.s. shape. As illustrated in Fig.~\ref{fig:Ne20_sp}, given the largest contribution of the $0(8\ 0)$ shape to the ground state, the largest contribution to the response peak indeed arises from its 2\hw~one-particle-one-hole excitation, that is $\omega = 2(10\ 0)$ dynamic deformation that corresponds to the GR-type vibrations within the $0(8\ 0)$ shape (see Fig. \ref{fig:shape}). However, there is a competing contribution to these $0^+$ states from the $\sigma = 2(10\ 0)$ shape with an equilibrium configuration already at $2\hw$ (that is, it does not allow $2\hw$ one-particle-one-hole de-excitations). Interestingly, the $2\hbar\Omega$ GR-type excitations of the $2(10\ 0)$ shape result in a dominant $4\hw$ dynamical deformation, $\omega= 4(12\ 0)$, seen as a peak at higher energies.

\begin{figure}[th]
\begin{center}
\begin{tabular}{c}
\includegraphics[width=0.45\textwidth]{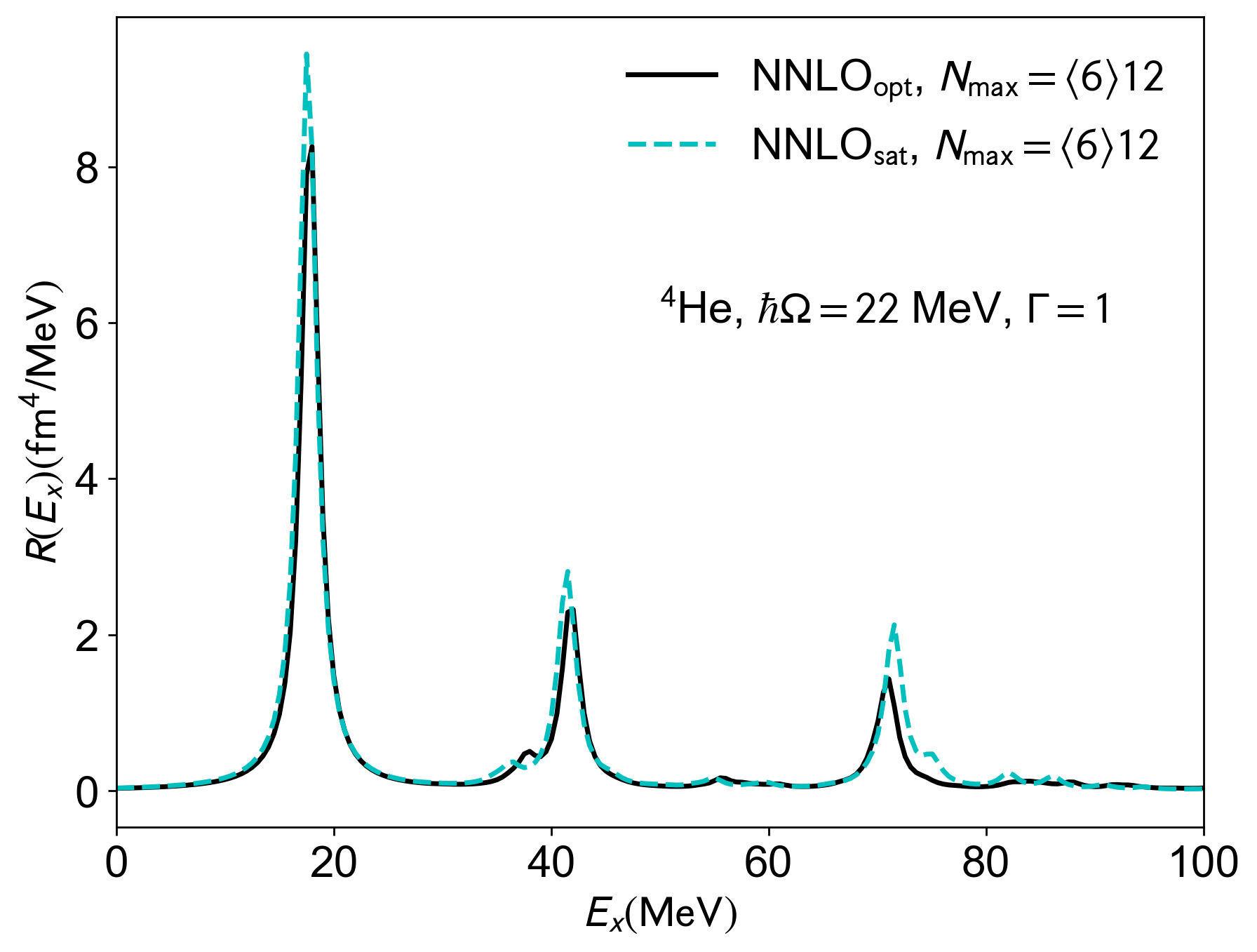} 	\\	\includegraphics[width=0.47\textwidth]{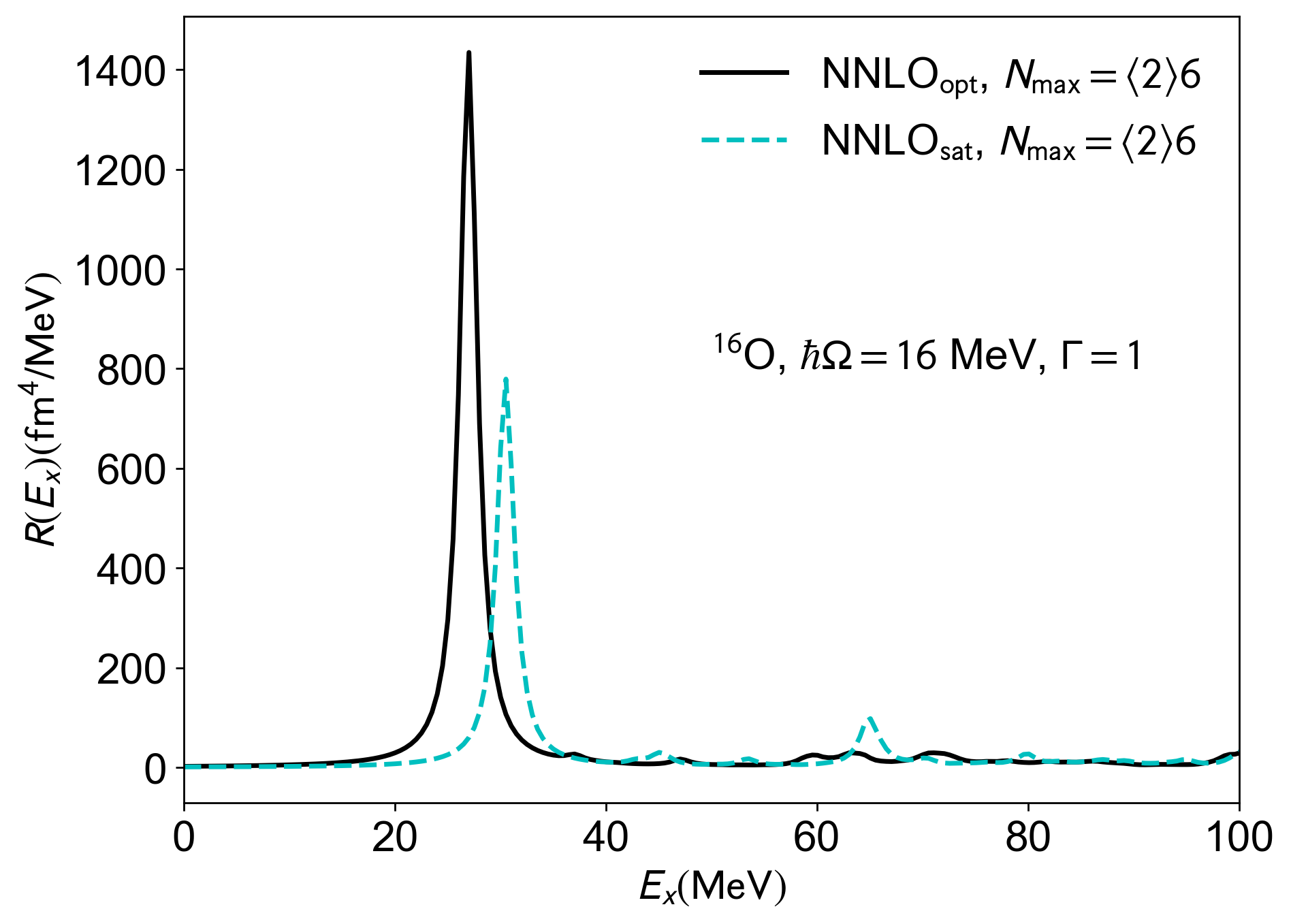}
\end{tabular}
\caption{Comparison of the monopole response functions for NNLO$_{\mathrm{opt}}$ and NNLO$_{\mathrm{sat}}$ for (a) $^{4}$He and (b) $^{16}$O. All response functions are calculated with $\Gamma=1$ MeV.}
\label{fig:sat_resp}
\end{center}
\end{figure}

In addition, we calculate response functions using a different realistic interaction, namely the NNLO$_{\mathrm{sat}}$ chiral potential with the NNN forces included as averages. We note that, in these calculations, the NNN forces are included as a mass-dependent monopole interaction \cite{LauneyDD12}, which  has an effect on binding energies, and, for example, for the $^{16}$O ground-state energy,  
the  7-shell NNN contribution is 20.46 MeV, resulting in $-127.97$ MeV total energy for $N_{\rm max}=8$ and \hw=16 MeV, which agrees with the experimental value of $-127.62$ MeV.
In Fig.~\ref{fig:sat_resp}, the monopole response functions for the closed-shell nuclei $^4$He and $^{16}$O are shown, comparing the results for both NNLO$_\mathrm{opt}$ and NNLO$_{\mathrm{sat}}$. Notably, the $^4$He response is similar for both interactions, with only slight differences in the heights of the peaks. In contrast, the main peak in the $^{16}$O response varies in the magnitude of the response and shifts slightly between the two interactions. This suggests that the  NNLO$_{\mathrm{sat}}$ NN+NNN(0) is slightly less compressible compared to the NNLO$_\mathrm{opt}$ and yields a smaller monopole sum rule, although the role of the remaining NNN forces, should be further investigated. Interestingly, the peak energy for $^{16}$O closely agrees with the one obtained in Ref. \cite{Stumpf2017} that uses NNN for both EM NN+NNN(400) \cite{EntemM03} and NNLO$_{\mathrm{sat}}$, although the broad and fragmented peak seen in Ref. \cite{Stumpf2017} may be a result of the interaction renormalization used.

\subsection{Nuclear compressibility}
\label{sec:K}

As mentioned above, breathing modes (giant monopole resonances) are expected to be observed at a high excitation energy as a result of almost incompressible nuclear matter \cite{Rowe16}.
This suggests, that with the ability to calculate response functions and examine giant resonances, we can utilize their connection to nuclear compressibility. The second-order derivative of the binding energy per particle gives us information about the stiffness of nuclear matter against variations in the density \cite{Harakeh2001}. This defines a compression modulus for infinite nuclear matter 
\begin{eqnarray}
K_{\infty} = \left. k^2_F \frac{d^2 (E/A)}{dk^2_F}   \right|_{k_F = k_{F_0}}, 
\end{eqnarray}
where $k_F$ is the Fermi momentum and $k_{F_0}$  corresponds to the saturation density for which the binding energy reaches its minimum. $K_{\infty}$ is often extracted through calculations of finite nuclei of mass $A$, denoted as $K_{A}$.
\begin{table}[t]
    \centering
    \begin{tabular}{|c|c|c|c|}
    \hline
         Nucleus &  $E_{GMR}$ [MeV] & $r_{\rm rms}^2$ [fm$^2$] & $K_A$ [MeV] \\
         \hline
         $^{4}$He & 27.0(1) & 2.05 & 36.0(2) \\
         $^{16}$O &  24(1) & 5.51 & 75(7) \\
         $^{20}$Ne &  25(1) & 7.22 & 109(6) \\
         $^{40}$Ca &   27(2) & 9.70 & 170(10) \\
    \hline
    \end{tabular}
    \caption{GMR centroid energies $E_{GMR}$  and  ground-state rms matter radius $r^2_{\rm rms}$ calculated in the SA-NCSM with the NNLO$_{\mathrm{opt}}$ chiral potential (for the \hw~and model spaces reported in Figs. \ref{fig:He-4_resp}a  and \ref{fig:mono_resp}), along with the compressibility $K_A$ calculated using Eq. (\ref{eqn:KA_GMR}). Error estimates are based on 10\% $\hbar\Omega$ variation. For comparison, the experimentally deduced centroid energies for $^{16}$O is $21.13(49)$ MeV \cite{Lui2001} and for $^{40}$Ca is $18.9(4)$ MeV \cite{PhysRevC.55.2811}. 
    }
    \label{tab:obs}
\end{table}

A widely used approach to estimate $K_{\infty}$ is called the microscopic approach \cite{Blaizot1995}. It starts with a microscopic nuclear interaction and typically utilizes an infinite nuclear matter approach that calculates the binding energy per nucleon as a function of the Fermi momentum \cite{Brockmann1984}. Recently, chiral potentials have been employed  (e.g., see Ref.~\cite{Drischler2019}) suggesting the need of repulsive NNN forces (e.g., see Ref.~\cite{PhysRevC.89.044321}).

In the so-called macroscopic approach, an estimate for the compressibility for finite nuclei is based on an empirical dependence on mass $A$, established earlier as a good first-order approximation. Namely, previous work \cite{Blaizot1980} has expressed the compressibility of a finite nucleus $K_A$ as the mass functional:
\begin{eqnarray}
\label{eqn:KA_lepto}
K_A &=& K_{\mathrm{vol}} + K_{\mathrm{surf}} A^{-1/3} + K_{\mathrm{Coul}} Z^2 A^{-4/3}\\
\nonumber
&+&  K_{\mathrm{sym}} \left (\frac{N-Z}{A} \right )^2 ,
\end{eqnarray}
where $K_{\mathrm{vol}}, K_{\mathrm{surf}}, K_{\mathrm{Coul}}, K_{\mathrm{sym}}$ are the volume, surface, Coulomb, and symmetry contributions to the compressibility of an $A$-body system and $N$ ($Z$) is the neutron (proton) number. In order to determine these coefficients, one possibility is to relate this expression for $K_A$ to the semi-empirical mass formula, however such approaches have been shown to be unreliable in previous work \cite{Treiner1981}. As such, one usually  connects $K_A$ to the centroid energy of the giant monopole resonance, $E_{\mathrm{GMR}}$, as
\begin{eqnarray}
\label{eqn:KA_GMR}
K_A = \frac{m_N}{\hbar^2} r_{\rm rms}^2 E^2_{\mathrm{GMR}}, 
\end{eqnarray}
where $m_N$ is the mass of the nucleon (in MeV) and
$r_{\rm rms}$ is the ground state rms matter radius
\cite{Blaizot1980}. 
In the macroscopic approach, one would find $K_A$ values for a variety of nuclei from Eq.~(\ref{eqn:KA_GMR}) and then use that data to fit the coefficients in Eq.~(\ref{eqn:KA_lepto}). In principle, $K_{\mathrm{vol}}$ can then be connected to $K_{\infty}$ \cite{Jennings1980, Blaizot1995}. However, a comprehensive analysis has shown that the values of $K_{\infty}$ estimated in this approach can have  large uncertainties \cite{Harakeh2001}.

Other approaches tend to combine some portion of the macroscopic and microscopic approaches in an attempt to reduce the amount of information one needs to calculate $K_{\infty}$, while also maintaining reasonable estimates for the uncertainty \cite{Garg2018}.

In this work, we combine the macroscopic approach of extrapolating to $K_{\infty}$ using Eq.~(\ref{eqn:KA_lepto}) with \textit{ab initio} calculations of $K_A$ using Eq.~(\ref{eqn:KA_GMR}), where $E_{\mathrm{GMR}}$, given in (\ref{eqn:GMR_SR}), and $r_{\rm rms}$ are calculated in the SA-NCSM framework. We take advantage of the SA-NCSM capability to provide accurate monopole sum rules (see Section \ref{sec:resp} and Ref. \cite{BakerLBND20}), and calculate the GMR centroid energies (\ref{eqn:GMR_SR}) for $Z=N$ nuclei in the range of $A=4$ to $40$ (Table \ref{tab:obs}). Together with the SA-NCSM calculations for the $r_{\rm rms}$ g.s. matter radii, an estimate for $K_A$ can be then given by Eq.~(\ref{eqn:KA_GMR}). This allows us to provide a fully microscopic estimate for $K_A$ and to  investigate  compressibility properties of the underlying chiral potentials directly through SA-NCSM structure and sum rules calculations. 

With the reported centroid energies for $^4$He, $^{16}$O, and $^{20}$Ne in Table \ref{tab:obs}, we  use Eq.~(\ref{eqn:KA_GMR}) to calculate $K_A$, the compressibility of these $A$-body systems. These results are shown in Table \ref{tab:obs} for NNLO$_{\mathrm{opt}}$. In the case of NNLO$_{\rm sat}$ NN+NNN(0), we report  $K_A=42.15 \pm 0.2 $ for $^{4}$He, 130 $\pm$ 18 for $^{16}$O, and 160 $\pm$ 25 for  $^{20}$Ne, yielding values that are systematically larger compared to the NNLO$_{\mathrm{opt}}$ case. This suggests that NNLO$_{\mathrm{opt}}$ is softer.

\begin{figure}[h]
\centering
\includegraphics[width=0.48\textwidth]{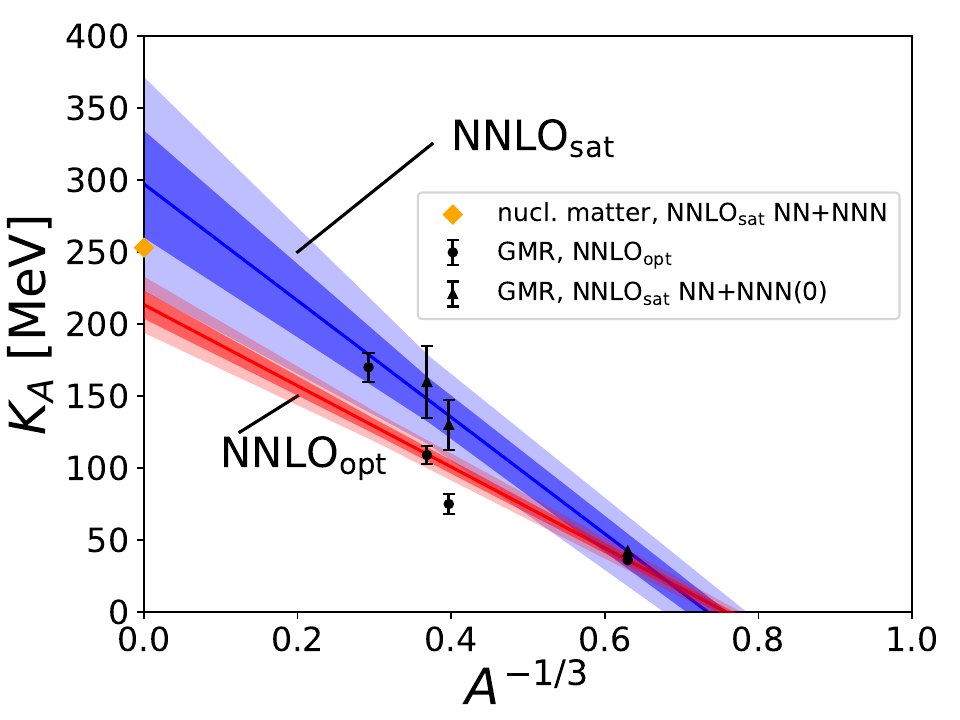}
\caption{Compressibility of an $A$-body system as a function of $A^{-1/3}$, as calculated by the SA-NCSM using the NNLO$_{\rm opt}$ (filled circles) and the NNLO$_{\rm sat}$ (filled triangles), with errors given by $\sim 10\%$ $\hbar\Omega$ variation. Also shown are the best fits for NNLO$_{\rm opt}$ (red solid line) and for NNLO$_{\rm sat}$ (blue solid line), with the corresponding 1$\sigma$ (darker shade) and $2\sigma$ (lighter shade) deviations based on an assumed Gaussian error distribution, as compared to a nuclear matter evaluation for NNLO$_{\rm sat}$ NN+NNN  \cite{Ekstrom2015} (orange filled diamond).}
\label{fig:nucl_comp}
\end{figure}

\begin{figure}[h]
\centering
\includegraphics[width=0.48\textwidth]{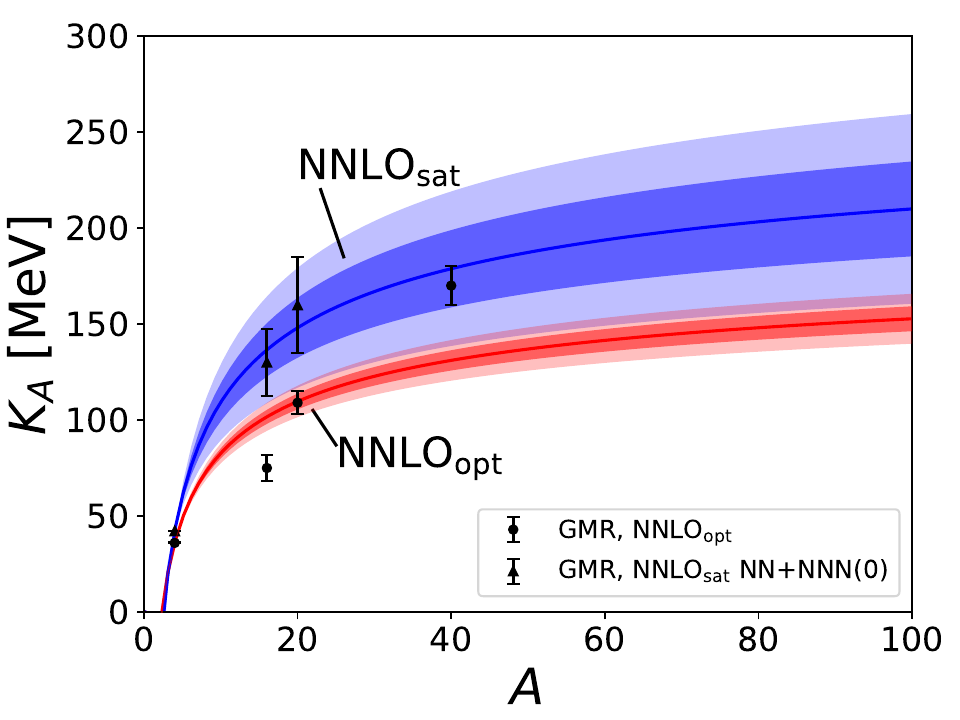}
\caption{Same as Fig. \ref{fig:nucl_comp}, but the compressibility coefficient is shown as a function of mass $A$.}
\label{fig:nucl_compA}
\end{figure}

To provide an approximate estimate for the infinite matter compressibility (Fig.~\ref{fig:nucl_comp}), we use the SA-NCSM outcomes for $K_A$ (Table \ref{tab:obs}) and a fit to Eq.~(\ref{eqn:KA_lepto}). As all of the data points involve symmetric nuclei ($N=Z$), the symmetry term  vanishes. Then we can introduce a new variable $x=A^{-1/3}$ and rewrite (\ref{eqn:KA_lepto}) as $K_A = K_{\mathrm{vol}} + K_{\mathrm{surf}} x$, given that the Coulomb term is smaller compared to the other two ($K_{\mathrm{Coul}} \sim -5$ MeV). We can thus use a linear fit to determine $K_{\mathrm{vol}}$ and $K_{\mathrm{surf}}$. A comparison between this linear fit, a linear fit with fixed values of $K_{\mathrm{Coul}}$, and a nonlinear fit is given in Table \ref{tab:fitparam}. Importantly, all of the fits yield the same value for $K_{\mathrm{vol}}$, which in turn determines $K_{\infty}$, whereas the other two parameters appear inconsequential, despite the fact that $K_{\mathrm{surf}}$ largely varies and both $K_{\mathrm{surf}}$ and $K_{\mathrm{Coul}}$ cannot be simultaneously constrained given the limited data set. Indeed, for \textit{ab initio} studies, reducing the number of parameters is important, since compute-intensive large-scale calculations can only provide a limited data set. We note that despite the reduced number of data points, the region of $A=15$ to $40$ appears significant in determining the best fit, which can be clearly seen in Fig.~\ref{fig:nucl_compA}, where the compressibility coefficient is plotted as a function of the mass number.  
\begin{table}[]
    \centering
    \begin{tabular}{|c|c|c|c|}
   \hline
Fit type	&	$K_{\mathrm{vol}}$	&	$K_{\mathrm{surf}}$	&	$K_{\mathrm{Coul}}$ \\
\hline
Nonlinear	&	$213.398$ & $	-380.599	$ & $	197.971 $\\
\hline 
Linear &	$213.398$ & $	-280.112	$ & $	-3.0 $\\
	($K_{\mathrm{Coul}}$ fixed) &	$213.398	$ & $	-279.112	$ & $	-5.0 $\\
	&	$213.398	$ & $	-277.612	$ & $	-8.0 $\\ 
 \hline 
Linear & $	213.398	$ & $	-281.613	$&	--   \\
   \hline
\end{tabular}
    \caption{Best-fit parameters for Eq. (\ref{eqn:KA_lepto}) as fitted to the $K_A$ values listed in Table \ref{tab:obs}. The fit and best-fit parameters used in this study to extrapolate to $K_{\infty}$ are presented in the last row. }
    \label{tab:fitparam}
\end{table}

The best linear fit yields a compressibility for infinite nuclear matter of $K_{\infty} = 213(10)$ MeV from the NNLO$_{\mathrm{opt}}$ chiral potential, which is a reasonable estimate given the very limited set of nuclei under consideration. This estimate is very close to the generally accepted value $250 < K_{\infty} < 315$ MeV \cite{Stone2014}, though it is on the lower side, pointing to the softness of the interaction as also mentioned above. For NNLO$_{\rm sat}$ NN+NNN(0), the best fit yields a compressibility for infinite symmetric nuclear matter of $K_{\infty} = 297(37)$ MeV. Within the linear regression uncertainties, this compares well with the compressibility  of 253 MeV 
for infinite  nuclear matter derived using the NNLO$_{\rm sat}$ NN+NNN reported in Ref. \cite{PhysRevC.91.051301} (see Fig. \ref{fig:nucl_comp}). This suggests that the properties of the breathing model can provide reasonable estimates for the nuclear matter incompressibility without the need for including (the full) NNN forces.
For comparison, Ref. \cite{Stumpf2017} includes the full NNN forces in NNLO$_{\rm sat}$ but uses renormalization in the nuclear medium; however, since $K_A$ is most sensitive to the GMR centroid energy $E_{\mathrm{GMR}}$, it is likely that the fragmented and broad GMR reported in Ref. \cite{Stumpf2017}  will yield larger $E_{\mathrm{GMR}}$. This suggests that the renormalization in the nuclear medium may result in even stiffer potential as compared to our estimate.

The approach described above represents a fully microscopic prescription to calculate $K_A$, and then extrapolate to $K_{\infty}$ using mass (density) dependent functional. It does not require infinite nuclear matter calculations, nor does it require experimental data. With the addition of more nuclear data, this approach holds promise to reduce uncertainties in $K_{\infty}$.

\section{Conclusions}
\label{sec:conclude}
We have presented \textit{ab initio} response functions from the Lanczos response method, using SA-NCSM wave functions as input. Specifically, we study response functions and sum rules for $^4$He, $^{16}$O, $^{20}$Ne, and $^{40}$Ca, where the SA-NCSM wave functions were calculated using the NNLO$_{\mathrm{opt}}$ NN and NNLO$_{\mathrm{sat}}$ NN+NNN(0) chiral potentials. We find that the $^4$He SA-NCSM response functions were in good agreement with response functions calculated using the standard NCSM, regardless of whether we were examining the monopole, dipole, or quadrupole transitions. We examined the giant monopole resonances and identified the underlying GR-type vibrations contributing to the giant resonance peaks, which are described by 2\hw~one-particle-one-hole excitations of the nuclear shape of the ground state. We also provided a fully microscopic estimates of the compressibility of a finite nucleus $K_A$, based on the GMR centroid energies $E_{\mathrm{GMR}}$ and g.s. rms matter radii. These are, in turn, used in a mass functional to extrapolate to the compressibility coefficient of  symmetric nuclear matter, as calculated from NNLO$_{\mathrm{opt}}$ and NNLO$_{\mathrm{sat}}$. Overall, these results indicate that the SA-NCSM is well-positioned to calculate response functions for open-shell nuclei and to explore collective and giant-resonance features of nuclei from first principles, as well as compressibility properties of the underlying chiral potentials. In doing so, our approach does not require infinite nuclear matter calculations, nor does it require experimental data. With the addition of more nuclear data,  it holds promise to reduce uncertainties in the $K_{\infty}$ compressibility coefficient of symmetric nuclear matter.

\begin{acknowledgments}
 This work was supported in part by the U.S. National Science Foundation (PHY-1913728, PHY-2209060), the U.S. Department of Energy (DE-SC0023532, DE-FG02-93ER40756), by the Deutsche Forschungsgemeinschaft DFG through the Cluster of Excellence [Precision Physics, Fundamental Interactions and Structure of Matter (PRISMA+ EXC 2118/1)]
  and by the Czech Science Foundation (22-14497S). This work benefitted from high performance computational resources provided by LSU (www.hpc.lsu.edu), the National Energy Research Scientific Computing Center (NERSC), a U.S. Department of Energy Office of Science User Facility at Lawrence Berkeley National Laboratory operated under Contract No. DE-AC02-05CH11231, as well as the Frontera computing project at the Texas Advanced Computing Center, made possible by National Science Foundation award OAC-1818253. 
\end{acknowledgments}

\bibliography{Combined}

\end{document}